# A TUTORIAL ON GRAPH THEORY FOR BRAIN SIGNAL PROCESSING

**Prof. Nikolaos Laskaris**
Department of Informatics,
Aristotle University of
Thessaloniki, 54124
Thessaloniki, Greece.

**Dr. Dimitrios A. Adamos**
Department of Computing,
Imperial College London,
London SW7 2AZ, U.K.

**Prof. Anastasios Bezerianos**
N.1 Institute for Health,
National University of
Singapore, 117456, Singapore

## ABSTRACT

This tutorial paper refers to the use of graph-theoretic concepts for analyzing brain signals. For didactic purposes it splits into two parts: "theory" and "application". In the first part, we commence by introducing some basic elements from graph theory and stemming algorithmic tools, which can be employed for data-analytic purposes. Next, we describe how these concepts are adapted for handling evolving connectivity and gaining insights into network reorganization. Finally, the notion of signals residing on a given graph is introduced and elements from the emerging field of *graph signal processing* (GSP) are provided. The second part serves as a pragmatic demonstration of the tools and techniques described earlier. It is based on analyzing a multi-trial dataset containing single-trial responses from a visual ERP paradigm. The paper ends with a brief outline of the most recent trends in graph theory that are about to shape brain signal processing in the near future and a more general discussion on the relevance of graph-theoretic methodologies for analyzing continuous-mode neural recordings.

## 1 INTRODUCTION

There are many reasons why graph theory is relevant to analyzing neural signals. The most obvious is that these signals concern the brain, the most complex living system in the universe. The approach of complex systems, mathematically founded over rigorous graph theory, appears nowadays as the most promising way in systems neuroscience [1–7] while it starts to gain popularity among neuroengineers as well [8]. From the early drawings of Ramón y Cajal until the recent thorough wiring diagrams emerging via *brain connectomics* research, the descriptive power of graph theory for handling and portraying relations has been repeatedly demonstrated. Apart from naturally encapsulating neural interactions and consequently conceptualizing brain mechanisms, the interest in graph theory is fueled by the potential of innovative information-mining techniques that continuously emanate within the new technological era of interconnectness and Big Data analytics. The term *"brain-web"*, coined by F. Varela [9], reflects the analogies between neuroinformatics and network-science and points to a wealth of data analytics, mostly of graph-theoretic orientation, that holds promise for making sense out of voluminous neural data. Finally, *graph signal processing* (GSP) constitutes an emerging field [10, 11], with novel techniques that are well suited for treating signals from sensor arrays of a given (often irregular) topology, as it is often the case in multisite recordings.

Cognitive neuroscience aims at elucidating the inherent mechanisms that underlie information processing in the brain and support various mental faculties. The rapid development of neuroimaging techniques enables the recording of task-relevant brain activity in a multitude of forms. The traditional approach seeks distinctive patterns of brain activation as neural correlates of cognitive (sub-)processes, by associating them with particular stimuli and behavioral responses. However, cognition is thought to emerge through integrated activities of neuronal populations throughout the brain, following the principles of segregation and integration [12]. Hence, the concept of functional/effective connectivity has become central for understanding the synergistic behavior of brain regions, which form distributed and partially overlapped networks, during a task [13, 14]. Thanks to the wide availability of relevant signal-analytic methodologies, connectivity can be assessed in various ways. Most often, connectivity measurements are turned into graph topologies, which undergo network characterization so as to associate them with functional brain states. More recently, the dynamic reconfiguration of brain networks has been brought under closer examination [15, 16]. Tracking the human brain's self-organization based on fMRI timeseries, when subjects are at rest or engaged in a given task, has become a very popular topic in current research [17, 18]. A similar trend is noticed for faster modalities, like EEG/MEG, where multisite recordings are exploited to track the networked brain on a millisecond basis [19, 20]. This kind of investigations requires efficient processing algorithms, suitable for handling a significant number of graphs (that ranges from few hundreds to thousands in the case of standard multi-trial recordings and grows even higher when group analysis is pursued at the level of single-trial dynamics) [21, 22]. The developed brain signal processing methodologies should comply with the notion of dynamic complex networks and, when employed in real-time applications (like Brain Computer Interfacing), be able to cope with streaming graph-data.

This paper concerns the use of graph-theoretic concepts for analyzing brain signals. It presents an overview of various algorithms and techniques that have been deployed in our research during the last two decades. For didactic purposes the paper splits into main parts. The first part articulates the appropriate theoretical background. It starts by providing the necessary mathematical foundation and discussing data-analytic tools for static graphs. It then proceeds with the adaptation of these tools for handling evolving graph-connectivity and ends with the concept of graph signals and their graph-spectral content. The second part includes representative applications and serves as demonstration of the preceding graph-theoretic tools. To ease presentation, we demonstrate our graph-analytic framework using multichannel EEG data from an event-related paradigm, in which single-trial ERP responses were recorded during an experiment that elicited the standard N70-P100 pattern. The paper concludes with future research directions and is accompanied by MATLAB code[1].

[1] Available on our website: https://neuroinformatics.gr

# 2 BASIC GRAPH-THEORETIC CONCEPTS

## 2.1 A GENTLE INTRODUCTION

Since its birth, with Euler's celebrated seven bridges of Königsberg problem, applied graph theory is characterized by imposing a level of abstraction to the problem at hand and emphasizing the relations among the involved entities [23]. Both concepts are easily exemplified in neuroscience, by considering recording sites (sensors of demarcated brain regions) and correlation matrices correspondingly. Herein, we introduce the basic notations, necessary for the remaining of the paper, with an example of synthetic data (see Fig.1), which can be thought of as corresponding to a spatial network.

A graph **G** is defined as a set of vertices (or nodes) $\boldsymbol{V}$ and a set of edges $\boldsymbol{E} \subset \boldsymbol{V} \times \boldsymbol{V}$ connecting the vertices. A simple example of a graph with N=10 vertices is provided in Fig.1a. It has been formed by selecting randomly 10 points $\{X_i\}_{i=1:N}$ on the plane and connecting with a line every possible pair, leading to a fully-connected graph. By assigning the pairwise distances as weights to the edges, the graph takes a tabular representation that coincides with the [N×N] Distance matrix **D**, where $D(i,j)= \|X_i-X_j\|_{L2}$. The image shown beneath the original graph visualizes the inter-point distance pattern and reflects the fact that the geometrical relationships are both reflexive and symmetric, i.e. D(i,i)=0 and D(i,j)=D(j,i).

A sparse connectivity graph $\mathbf{G_1}$ can be defined by keeping only the links shorter than a given threshold and transforming the previous weights D(i,j) to connectivity strengths $W(i,j) = \max_{ij}\{D(i,j)\} - D(i,j)$ . The derived graph is shown in Fig.1b, along with its tabular representation **W** usually referred to as **Weighted Adjacency matrix**. This format is the most common in current network-neuroscience studies, since any type of functional connectivity measures can be laid in this way. The node strength, defined as $s_i = \frac{1}{N-1}\sum_j W(i,j)$, provides the simplest way to characterize the importance of a node in the network. Nodes of high strength are 'central' in the underlying communication pattern (*Centrality* is an important concept in complex network analysis. There are various node and edge centrality metrics [24], with each one targeting a particular aspect of the underlying connectivity pattern). In the network visualization step, the node size is usually costumed to reflect this information.

A more detailed description of the structure encoded in graph $\mathbf{G_1}$ can be provided by means of shortest path derivation. As shown in Fig.1c, the shortest path between node-1 and node-7 is the sequence (1-9), (9-3), (3-7). It is the particular sequence of edges, consisting of touching edges, that ensures reaching node-7 from node-1, while traversing the smallest distance over the connectivity graph. There are well known algorithms for deriving, simultaneously, all shortest paths. The physical meaning of shortest paths is that they reflect how easily information flows, in a feed-forward manner, over a given network. The estimation of the shortest path lengths $l_{ij}$ is an integral part of important graph metrics like the *small-worldness* and the *Global efficiency* [3, 5]. The latter metric is a measure of integration defined as

$$GE = \frac{1}{N(N-1)} \sum_{j,i\neq j} \frac{1}{l_{ij}}$$

that reflects the efficiency of information exchange in a network in which all nodes are capable of concurrently exchanging information via shortest paths. A "localized" version of the above formula is used to define the measure of *Local efficiency*, which characterizes the information exchange in the subnetwork around each node. Besides these metrics, the shortest-path lengths (also known as *geodesic distances* in manifold-learning theory) are necessary for graph-embedding algorithms like the famous ISOMAP [25], a non-linear dimensionality reduction method (efficient in retrieving parsimonious representations from complex neural data [26] ), which is realized via classic *multidimensional scaling* (MDS) of the geodesic distances. The associated geodesic-distance matrix **GD** is of the same size as the original distance-matrix **D**, but its entries $GD(i,j)=l_{ij}$ have higher values for the node-pairs that there are no direct-link in the connectivity graph in $\mathbf{G_1}$ (for instance, GD(1,7) > D(1,7)).

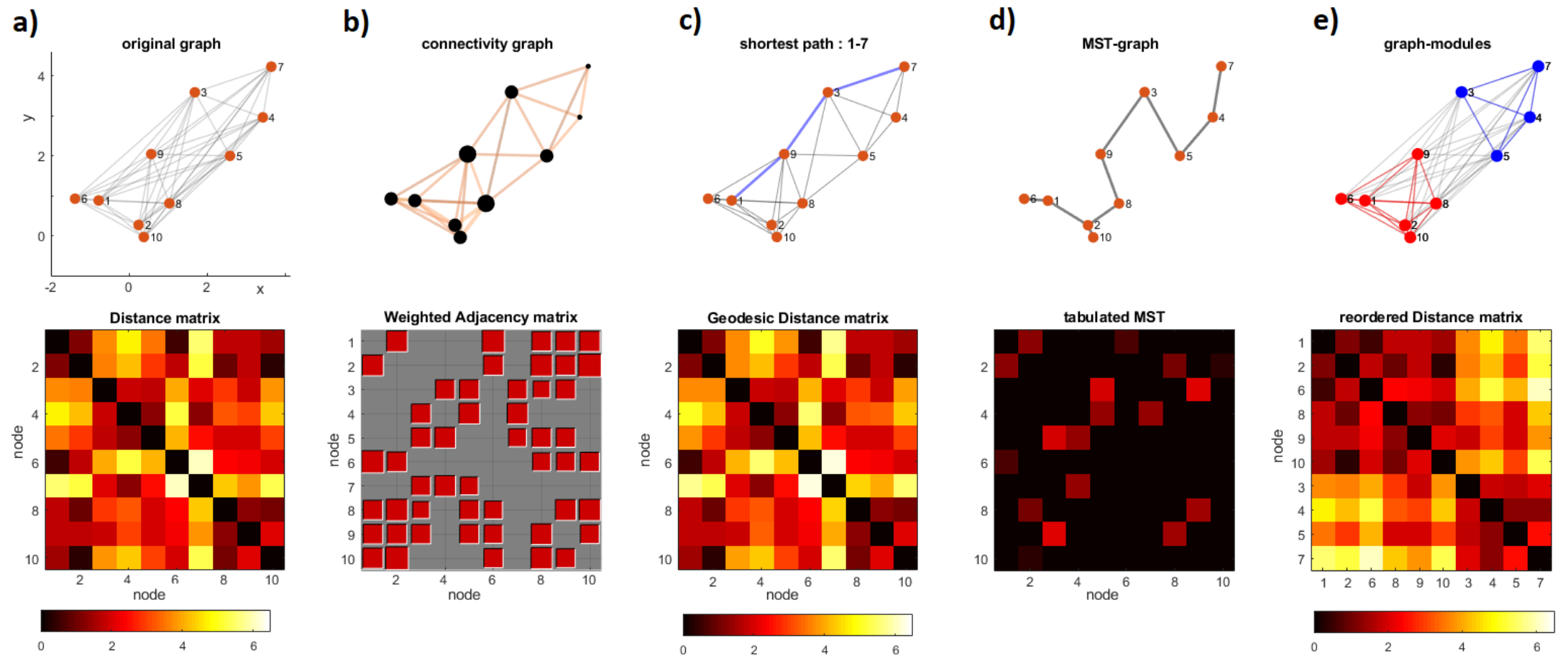

**Figure 1**. Using a synthetic connectivity graph, as derived within a 2D coordinate space, to illustrate basic graph-theoretic concepts. Top row refers to the graph and its transformations, while bottom row to their tabulated versions.

A reduced description of the connectivity structure in graph $\mathbf{G_1}$ can be readily derived through the computation of the minimum spanning tree (MST) graph, shown in Fig.1d. MST is among the sparsest connected graphs that can be derived from the original graph [27]. It is formed from the subset of N-1 edges that connect all nodes with the minimum wiring length. It is useful for ordering the nodes in the graph and defining topological descriptors like the diameter.

The last operation that is demonstrated in Figure 1 is this of modularity analysis or community detection [28, 29], a term that is used to describe algorithmic processes working towards graph-segmentation [30]. The two non-overlapping subgraphs, indicated in Fig.1e, have been demarcated using a standard spectral graph clustering step [31]. This step included the computation of *graph Laplacian* $\mathbf{L}=\mathbf{S}-\mathbf{W}$ (where $\mathbf{S}$ is a diagonal matrix containing the node strengths), the derivation of the principal eigenvector and its thresholding. In general, the scope of this kind of analysis is to group nodes into clusters (or communities), such that the similarity within clusters is high and the similarity between clusters is low. A *modularity-index* is used to quantify the quality of a given partition and it is very important to be compared against the result from the application of the same algorithm to randomized graphs. Among the available graph-clustering algorithms, in our research, we've given preference to *dominant-sets* algorithm [32–35] due to its computational efficiency and its descriptive power [36]. The communities (i.e. dominant sets) are detected iteratively, based on a given Weighted Adjacency matrix, and ranked according to the Cohesiveness Index that expresses the average mutual coherence within a detected community. Finally, the number of communities is a self-tuned parameter. The resulting structural description includes an ordered list of subgraphs along with their cohesiveness. The tabular representation of partitioned graph in Fig.1e reflects the imposed organization.

## 2.2 TIME-VARYING GRAPHS AND EVOLVING NETWORKS

Recently, the study of time-varying graphs has attracted an increasing interest [37]. Dynamic connectivity is currently brought under investigation in almost every scientific field dealing with streaming or longitudinal data. Time-resolved graph structure is considered ideal to track alterations and abrupt changes in adaptive complex networks [38]. Brain's self-organization and task-related re-organization, therefore, constitute fascinating application fields for this branch of modern graph-theory. Next, we discuss the two most direct ways to exploit the concepts developed in 1.1 for the purpose of monitoring graph structure.

Returning to the spatial network of Fig.1, we produce 20 different versions ${}^{t}\{X_i\}_{i=1:N}$ of the original point-sample. The first 10 versions were only slightly perturbed by adding small gaussian noise, while the last 10 versions were systematically modified so as to simulate the motion of a sub-network (the five top-most nodes) towards the right and down. The time-series ${}^{t}\mathbf{G}$, t=1,2,…,20 of the connectivity graphs is shown at the top of fig.2a, where time stamps have been associated with each graph. The associated Weighted-Adjacency matrices ${}^{t}\mathbf{W}$ are shown beneath these graphs.

The first way to detect this reconfiguration is to analyze each connectivity snapshot ${}^{t}\mathbf{W}$ independently by estimating a network metric and deriving a temporal signature of network organization like the ones shown in Fig.2b and Fig.2c. The second way is to define a suitable distance for comparing any two graphs dist(${}^{t1}\mathbf{G}$, ${}^{t2}\mathbf{G}$) and use it to derive a "projection" of the dynamics in a space of reduced dimensionality [39] (for instance via ISOMAP acting on the connectivity-graph of instantaneous graphs). This is exactly the concept behind the recently introduced methodology of *leading eigenvector dynamics* [40]. For our example, this methodology was adapted so as to derive the unidimensional signature of dynamics shown in Fig.2d. It comes in the form of a distance-profile, dist(${}^{1}\mathbf{G}$,${}^{t}\mathbf{G}$)=dist(${}^{1}\mathbf{W}$,${}^{t}\mathbf{W}$)=$\|{}^{1}V_1-{}^{t}V_1\|_{L2}$ , t=1,2,…,20, where ${}^{t}V_1$ was the eigenvector of ${}^{t}\mathbf{W}$ associated with the largest eigenvalue.

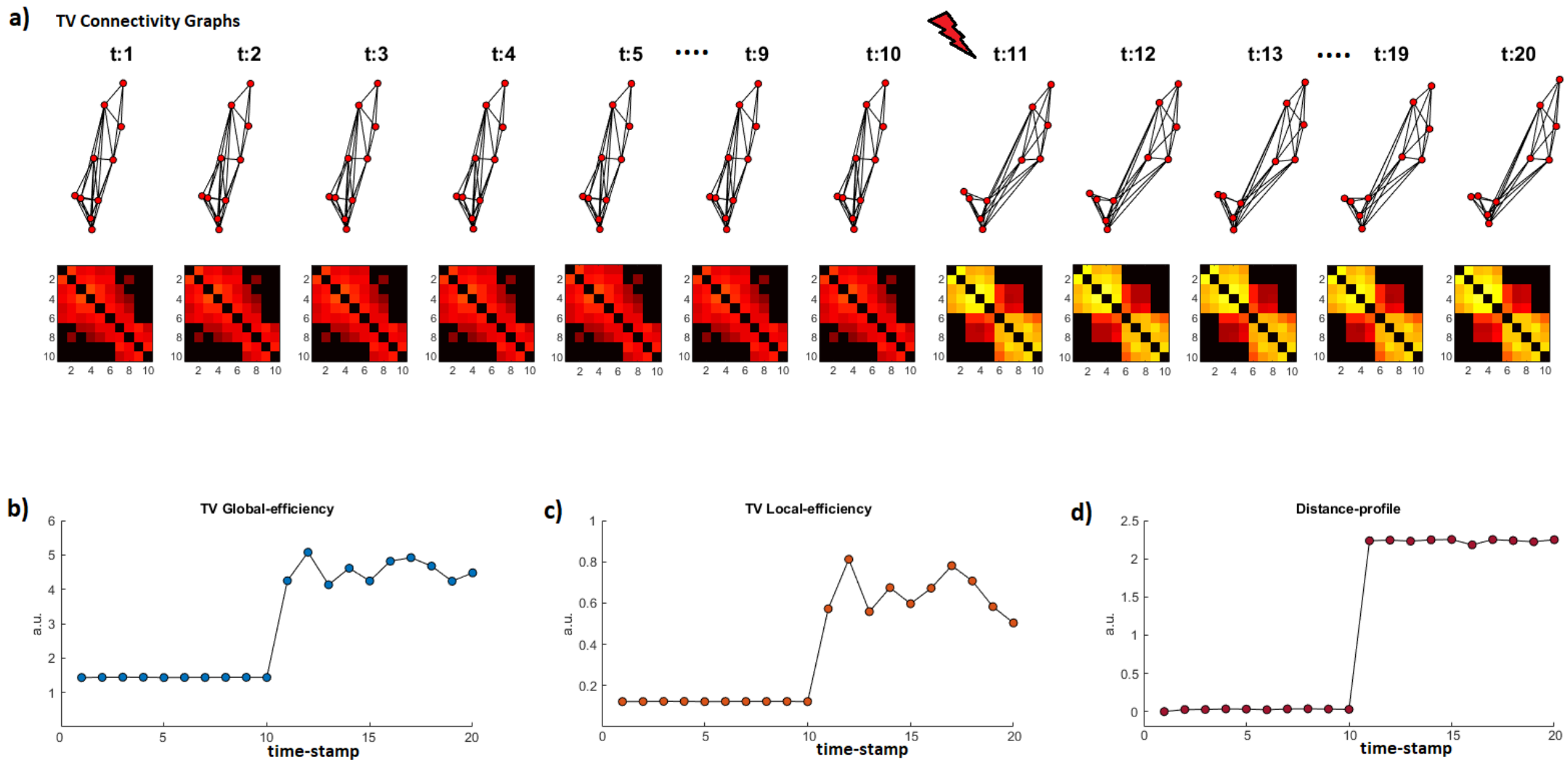

**Figure 2**. Studying dynamic connectivity graphs as means to describe an evolving network. a) Timeseries of connectivity graphs with their associated weighted adjacency matrices. b-c) Time-varying estimates of network metrics. d) Change point detection in network evolution.

## 2.3 Graph Signal Processing (GSP)

GSP combines spectral graph theory with standard signal processing techniques offering a unified framework to handle signals over irregular domains, as it is the case of electrical/magnetic field measurements recorded via a given sensor array in EEG/MEG and the concurrent brain BOLD traces from a predefined set of regions of interest (ROIs) [41–43]. It appears as the most suitable way to deal with multidimensional signals registered from a distributed network. Graph Fourier Transform (GFT) is one of the cornerstones of GSP. It transfers the potential of classical Fourier transform to the domain of graph-signals and can lead to robust signal analytics that respect the inherent constraints posed by the observation space.

To illustrate the concepts of a graph-signal and its GFT-spectrum, we have used the spatial network of Fig.1b to define the topology of the domain where our exemplar graph-signals reside. Figure 3a includes two graph signals. Each one is formally defined as a function that assigns real-valued vectors to the ordered set of nodes in graph $\mathbf{G}$, i.e. $\mathrm{X}_1, \mathrm{X}_1\colon \mathbf{V} \rightarrow \boldsymbol{R}^{\mathbf{10}}$. We used a common color-code to indicate these two signals on top of the network topology and make clear the distinction between a smooth and non-smooth graph signal (vectors $\mathrm{X}_1$ and $\mathrm{X}_2$ accordingly). In the former case, connected nodes are associated with similar signal-values. Since a smoothness-index can be formalized via graph-Laplacian $\mathbf{L}$, in the form of $\mathrm{X}_\mathrm{i}^\mathrm{T}\, \mathbf{L}\, \mathrm{X}_\mathrm{i}$ , the eigen-decomposition of $\mathbf{L}$ provides a set of orthonormal basis functions for representing the graph signals. By denoting the set of eigenvectors $\boldsymbol{U} = [\mathrm{U}_1, \mathrm{U}_2, \ldots, \mathrm{U}_{|\mathrm{V}|}]$ , after they have been ordered ascendingly according to their eigenvalues $\lambda_k$, we have at our disposal an operator that readily compares any graph signal against the graph-Laplacian eigenmodes. These eigenvectors have a role similar to that of complex exponentials signals, which are employed in standard discrete Fourier Transform for analyzing a time domain signal. As shown in Fig.3b, the eigenvectors associated with low eigenvalues (e.g. $\lambda_k = 1, 2$) display slow variations across the graph domain, while the ones related with larger eigenvalues (e.g. $\lambda_k = 9, 10$) display swift variations. GFT is derived by projecting the graph signals along the directions spanned by the Laplacian eigenvectors, i.e. $Z_i = \boldsymbol{GFT}(\mathrm{X}_\mathrm{i}) = \mathbf{U}^\mathbf{T}\mathrm{X}_\mathrm{i}$ , $i = 1{,}2$. The two associated patterns of signal energy distribution over the GFT-components, $GFT_Power_i(\lambda_k) = |Z_i(\lambda_k)|^2$, are shown in Fig.3c to clarify the distinction between a low-frequency and a high frequency graph signal.

GFT has already found application in fMRI studies of learning [41, 42], and more recently in crafting decoders for motor-imagery BCI [43]. Apart from its descriptive power, and due to the existence of inverse-GFT, it also is plays an instrumental role in designing graph-filters, which are matrix operators that enhance or attenuate particular GFT-components [44, 45].

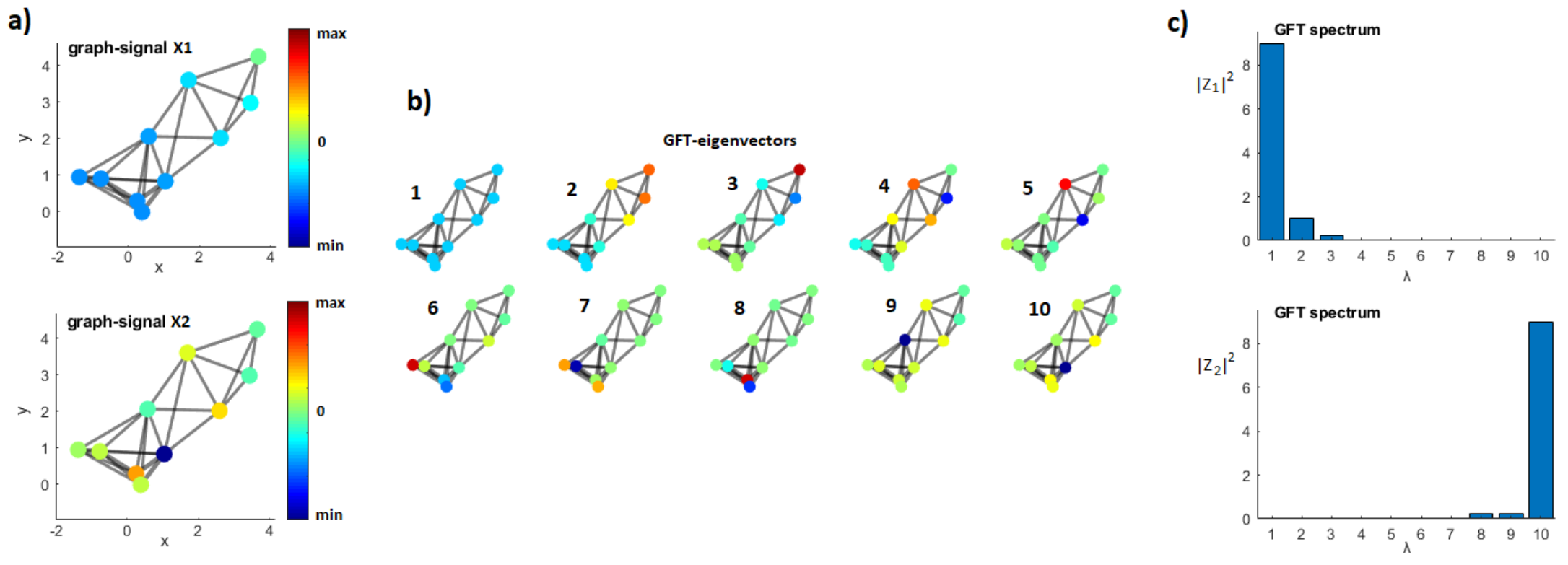

**Figure 3**: Exemplifying the spectral content of graph signals

# 3 GRAPH THEORY IN ACTION

To demonstrate the actual use of the presented methodologies, we utilized single-subject's data which had been part of a recent EEG study and concerned the challenging problem of differentiating between attentive and passive visual responses. A detailed description of the experimental paradigm, the recording procedure and the preprocessing of the data can be found therein [46]. In short, the subject participated in a pc-game, where he was engaged into the scenario of driving a racing-car by means of eye-saccades. While the car was moving ahead, a wall appeared suddenly on the left (or right) of the road (refer to Fig.4) and he had to avoid it by performing an anti-saccade (i.e. moving his gaze towards the opposite direction). The beginning of each trial was indicated by a fixation cross appearing in the center of the screen (baseline period). After a period of two seconds, a checkerboard-patterned wall appeared either on the left or right side of the screen, pseudo-randomly. Four seconds later, the central fixation cross disappeared, and the subject had to produce a saccade towards the opposite side of the checkerboard. Before the beginning of each consecutive trial there was a resting period lasting five seconds. Participant's brain activity was recorded at the sampling rate of 512 Hz, by means of a 64-channel EEG equipment. An additional recording session was also performed using the same stimuli, but in which the subject had been instructed not to perform the anti-saccade and passively view the stimuli on the screen. The first condition will be referred to as "attentive" condition, while the second as "passive". The time instant of checkerboard-pattern onset is denoted as 0-time. We note, here, that around 100ms after the stimulus onset a well-defined brain activation pattern, known as P100 response, emerges in the sensor-space that greatly differs between the "left" and "right" responses based on topographical laterality that builds over occipital and parietal brain areas. On the contrary, the differentiation between attentive and passive responses (when the wall appears on the same side) is a difficult pattern-classification task, when treated at a single-trial analysis level, and of great importance since it may be exploited in endogenous BCIs.

The following demonstration begins with the graph-theoretic analysis of passive visual responses when the wall appeared on the right side and finalizes with the contrast between the attentive and passive brain reactions happening within the first 500 milliseconds, with the scope of identifying brain state descriptors that can facilitate fast user-machine interaction. (Prototypical scripts for reproducing most of these results are available on our web site[2]).

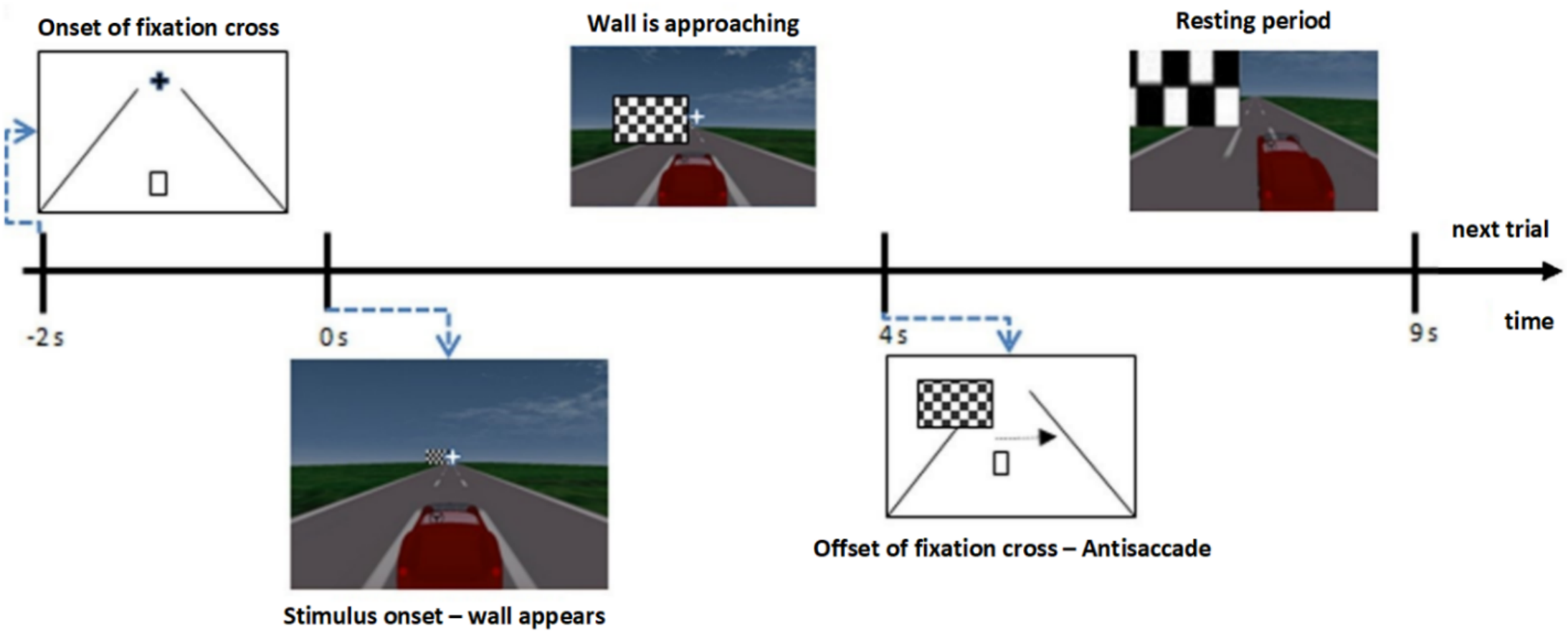

**Figure 4**: The timeline of a single-trial (ST). The pc-game had been developed by DIANA research group for the BRAINS project[3]

[2] https://neuroinformatics.gr
[3] http://www.diana.uma.es/brains/

## 3.1 Representing Averaged Response

Time-locked averaging of single-trial responses $x_i^j(t)$, j=1,2,…$N_{sensor}$, i=1,2…,$N_{trial}$ is an important initial step to get a gross picture of brain's reaction to a particular stimulus. The averaged signal $x_{ave}^j(t) = {}^1/_{N_{trial}} \sum_i x_i^j(t)$ is visualized in the form of successive topographies in an attempt to identify the sensors (and approximately the cortical locations) of maximal deflections. An alternative graph-theoretic approach is shown in Figure 5. It incorporates the semantic mapping of multichannel averaged response within a reduced space and has been derived via ISOMAP with an algorithmic procedure summarized within the following steps [47–49]. An ($N_{sensor} \times$ p) data-matrix $\mathbf{X_{ave}}$ was first formed, with $i^{th}$-row containing the corresponding segment of the averaged response, i.e. $X_{ave}\,(i,:) = [x_{ave}^j(t_1), \dots, x_{ave}^j(t_2)]$, where the limits mark the end and beginning of the latencies of interest. These segments were treated as points in a p-dimensional coordinate space, and a connectivity graph was then formed (based on "ε-ball" rule with a radius equal to average inter-point distance). Next, this graph was used to define geodesic distances in the space of response-variation. Finally, MDS technique was employed for the embedding of multidimensional-points within a plane. The end-result is a 2D scatter-plot diagram, where each point represents a sensor and their inter-point geometrical relationships reflect how the averaged response varies as we move between sensors.

Figure 5a includes the sensor-array as a spatial-network, with graph connectivity reflecting the closeness between sensors. Based on their positions, the sensors have been grouped into three groups: left/right hemisphere and midline. A common colormap was derived (by merging three distinct colormaps) and the nodes have been colored in a way that the imposed hue variations clearly indicate the group and the position within the group. This color-code is preserved in the butterfly plot shown in Fig.5b, and in the associated plot of signal-to-noise-ratio (SNR) profiles (An SNR-estimator for multi-trial responses was used, which assumed that noise was additive and independent across trials [49] - Fig.5c). The ISOMAP-based "projection" of the sensor-array, based on the similarity in temporal patterning, is shown in the Fig.5d. On top of this point-sample image, the original (spatial) connectivity pattern of the sensor-array (initially shown in Fig.5a) has been sketched, providing a vivid picture of how a stimulus may naturally lead to the reorganization of local brain activations. Considering, jointly, the deflections seen in the butterfly plot and their ISOMAP representation, one can clearly infer a sequence of stimulus induced events. There is a first group of intensified "yellow brainwaves" peaking at 100 milliseconds around PO7 sensor, followed by a second group of moderately elevated "green brainwaves" formed around P8 sensor. The sequence ends with a later widespread activation, of much lower SNR, that includes sensors located frontally (F8, AF8 and AF7).

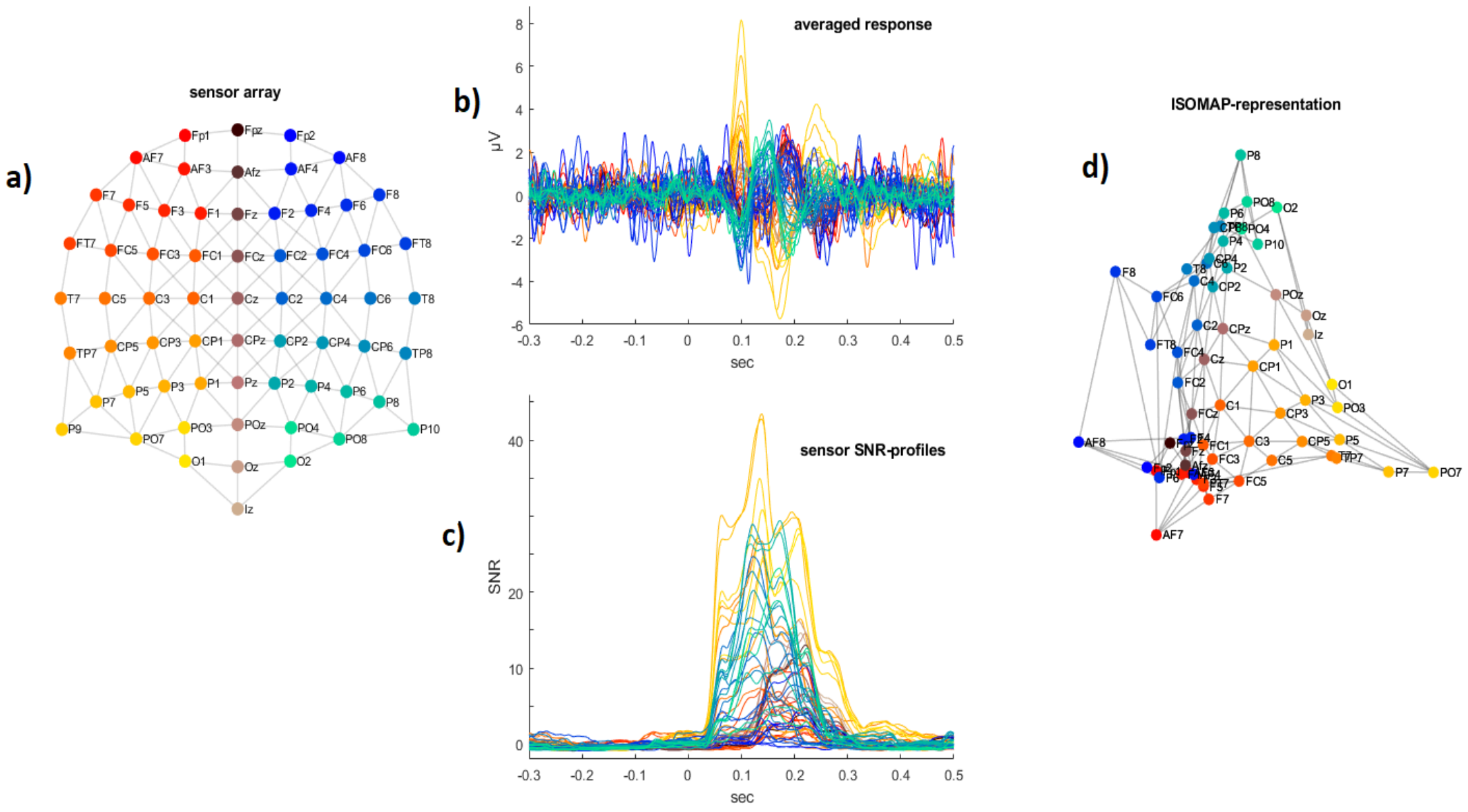

**Figure 5**: ISOMAP representation of visual evoked response variation. a) Topographical arrangement of sensors. b) The butterfly plot of averaged response; traces have been colored according to the colors attached to the sensors. c) The sensor SNR-profiles colored accordingly. d) A reduced geometric

representation of the differences between sensors based on their temporal patterning in averaged response within the first post-stimulus interval of 300 msec.

## 3.2 Visualizing Single-Trial Response Variability

The variability of brain responses, when treated at the level of single-trials, has been under investigation in a series of studies (e.g. [50–52]). Its characterization remains an open problem, since it poses serious restrictions in understanding the real-time information processing within the brain and the development of BCI and neurofeedback applications. In the following, we present a graph-theoretic approach to this problem that is based on MST-graph and intends to organize the single-trial responses in an intelligibly way [53]. The MST-based approach can be used to quantify the extend of variability, contrast it between recoding conditions and design selective-averaging algorithms for the recovery of response (i.e. in place of ensemble averaging). The MST-ordering of single-trials is demonstrated in Fig.6 based on the traces from the PO7 sensor, which was the one associated with the highest SNR and hence providing a clear picture of P100 response and its associated variability. An ($N_{\text{trial}} \times$ p) data-matrix $\mathbf{X}^{\mathbf{PO7}}$ was first formed, with $i^{\text{th}}$-row containing the corresponding segment of the single-trial response, i.e. $X^{PO7}(i,:) = [x_i^{PO7}(t_1), \ldots, x_i^{PO7}(t_2)]$, where the limits were defined, based on the SNR profile of the sensor, so as to include the latency range (50-300) msec. These segments were treated as points in a p-dimensional coordinate space, and the MST-graph was formed next. By means of an associate algorithmic procedure [53], known as "MST-planing", it was finally represented within a plane as shown Fig.6a. Each node therein corresponds to a single-trial response, and the extend of variability is reflected in the bimodality of the edge lengths, or alternatively in the *eigenvector centrality* scores assigned to the nodes. The majority of single-trials form a tight core, while the rest of them appeared as randomly scattered. To gain some insights into the form of trial-to-trial variability some trials have been selected, according to their placement on the MST-graph, and their waveforms are stack-plotted in Fig.6c. Apparently, the single-trial responses from the core of the distribution (e.g. 39 and 42) align better with the standard patterning seen in averaged N100 VEP response. On the contrary, the outliers (e.g. the remaining labeled trials) show a deviant patterning with the common characteristic of increased signal amplitudes starting even from the pre-stimulus interval. It should be mentioned here that the preprocessing of the single-trial data incorporated a standard data-cleaning procedure (removal of bad sensors/trials and ICA-denoising) for removing artifactual signals. Therefore, it is safe to assume that the visualized variability cannot be attributed to biological noise.

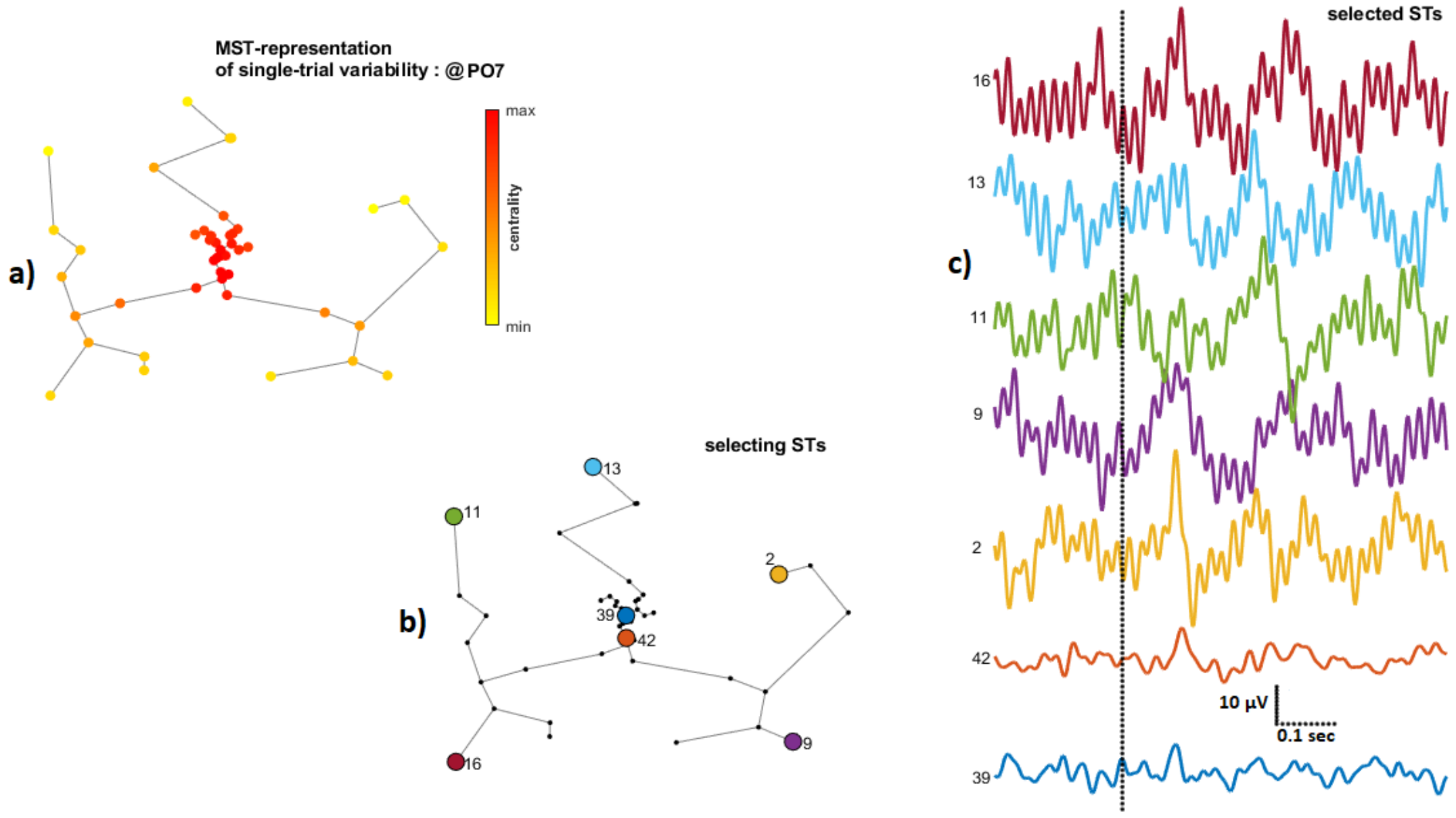

**Figure 6:** MST-representation of single-trial (ST) variability of the visual evoked response registered at the sensor of highest SNR, PO7.

## 3.3 Mining snapshots of connectivity from the evolving brain network.

The monitoring of cognitive state is of paramount importance in neuro-engineering [78] and is currently pursued by decoding dynamic functional connectivity from subjects at rest or engaged in mental tasks. In particularly, the case of event-related brain's functional reorganization lies at the junction of time-varying graphs and brain signal processing and offers a unique opportunity to unravel brain's mechanisms and associate them with well-known signatures of brain function, like ERP components [54–58].

In this section, we exemplify the mining of dynamic connectivity graphs from the multichannel visual responses as a means to summarize stimulus-induced network re-organization. The employed algorithmic procedure started by band-pass filtering the single-trial signals and then deriving the instantaneous phases via Hilbert transform. These phases $\theta_j^m(t), m = 1,2 \dots, N_{sensor}, j = 1,2 \dots, N_{trial}$ were utilized by a phase synchrony estimator known as Phase Locking Value (PLV) [59], in a time-resolved manner [60], to derive temporal sequences of pairwise connectivity strengths

$$PLV_{lm}(t) = \left| \frac{1}{N_{trial}(2w+1)} \sum_{t'=t-w}^{t+w} \sum_{j=1}^{N_{trial}} e^{i\,\Delta\theta(t')} \right|$$

where, $\Delta\theta_j(t) = \theta_j^l(t) - \theta_j^m(t), m, l = 1,2 \dots, N_{sensor}, j = 1,2 \dots, N_{trial}$ and w a resolution parameter that controls the temporal window. The derived 3D-tensor represented a timeseries of connectivity graphs $^{t}\mathbf{G}$ spanning the sensor-array. The timeseries of associated weighted-adjacency matrices $^{t}\mathbf{W}$ were such that $^{t}W(l,m)=PLV_{lm}(t)$. By averaging the connectivity snapshots from the pre-stimulus interval, a "baseline" connectivity pattern was first derived and then used to form the following distance profile $Profile(t)=dist(^{baseline}\mathbf{W}, {^{t}\mathbf{W}})=\|^{baseline}V-{^{t}V}\|_{L2}$, where $^{t}V$ is the leading eigenvector of $^{t}\mathbf{W}$. The derived signal, shown in Fig.7 for two different brain rhythms, quantified the extend of stimulus evoked changes in functional organization and used to select some instances of network organization that were analyzed in terms of graph-theory and presented comparatively to unravel the dynamic and distributed character of visual response. The mined connectivity patterns were first trimmed (to reduce clutter), using a common threshold at PLV-level of 0.7, and then overlaid over the sensor-array. Additionally, a centrality score (computed before trimming) was coded as node size. An interesting observation is that the evolving network appears spatially imbalanced, with its more "active" nodes located occipitally and contralaterally to the stimulus (i.e. the communication hubs coincide with the locations of maximal P100 response). Apart from this, the α-rhythm network is more widespread than the β-rhythm network and complies with the existence of a fronto-parietal subnetwork. On the other hand, the β-rhythm network seems to engage faster the frontal brain areas in the process of stimulus-perception than its α-rhythm counterpart.

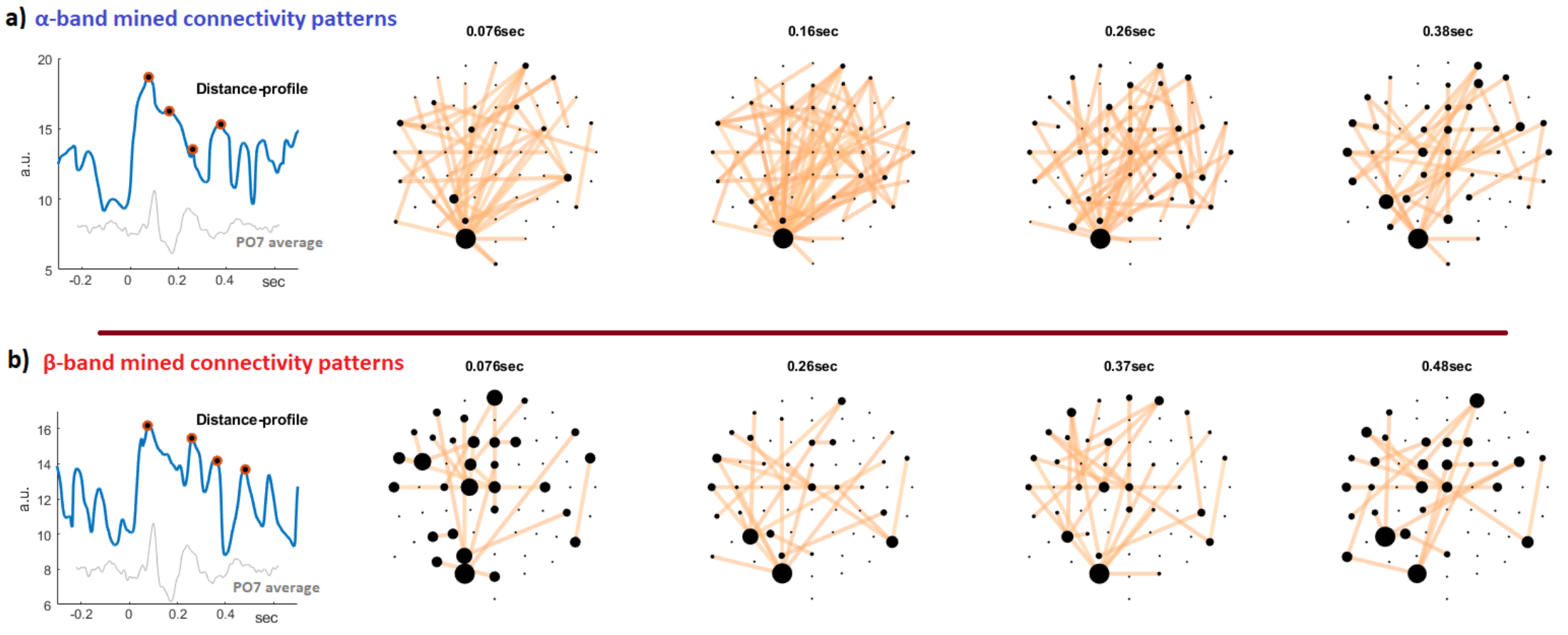

**Figure 7**: Monitoring stimulus-induced changes in brain network organization and picking "energetic" instances to provide a topological characterization by means of the descriptor of page-rank centrality. The selected connectivity patterns have been presented as connectivity graphs, trimmed at a common

threshold of connectivity strength, and the centrality measure has been indicated by the node size. Top/bottom row refers to time-varying graphs estimated from signals filtered within α-band/β-band.

We, next, continue with the characterization of these mined connectivity patterns, by referring to a graph-theoretic approach that adopts a perspective that is considered orthogonal to the analysis involved in the above-mentioned results and, hence, provides additional or complementary information. While Figure 7 highlights the important nodes from the overall communication pattern and therefore puts emphasis on the integration of the functional brain network, Figure 8 examines the segregation tendencies within the network by comparing the modular structure detected from each connectivity pattern based on *dominant-sets* algorithm [35]. We only show, in ranked fashion (red>green>blue), the communities that were more coherent than the ones detected from connectivity patterns in the pre-stimulus interval. The stimulus induced repertory of these communities appears rather restricted and includes modules that can be mostly associated to the underling visual brain areas. Interestingly, the most coherent functional group emerges at the latency of 160 milliseconds for the α-rhythm network and is contralaterally to the stimulus location in the visual field. On the contrary, a well-formed community emerges within β-rhythm network earlier, at the latency of 76 milliseconds, and locates ipsilaterally.

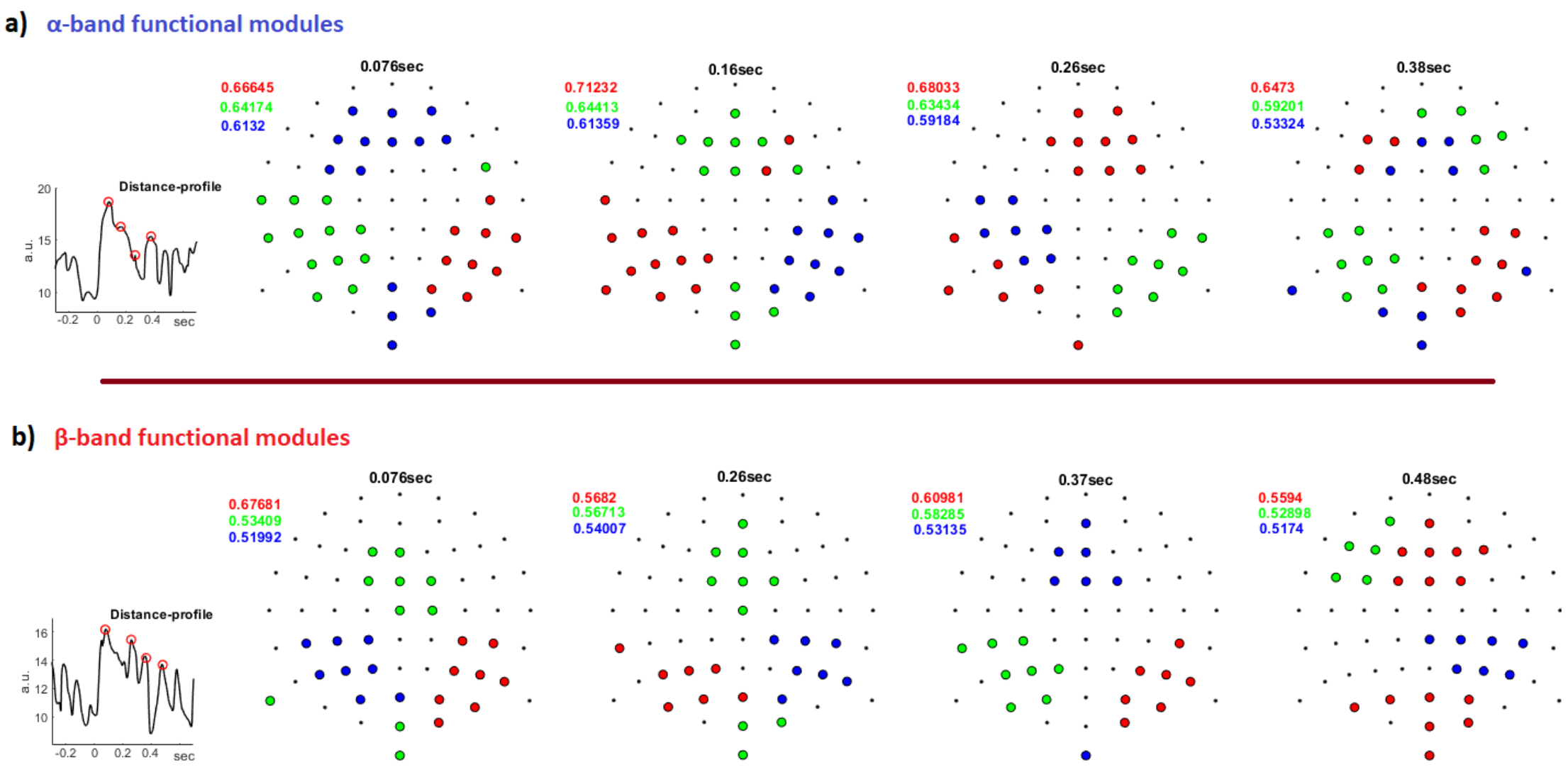

**Figure 8**: Monitoring stimulus-induced changes in brain's network organization and picking "energetic" instances to perform modularity analysis. Only "significant" communities are shown together with their cohesiveness level. Top/bottom row refers to time-varying graphs estimated from signals filtered within α-band/β-band.

## 3.4 GFT-steered enhancement of Evoked Response

The final example from the analysis of passive visual responses demonstrates a novel, data-driven, GSP technique aiming at signal enhancement. In a nutshell, the multichannel signal is first Graph-Fourier-Transformed [43], the individual projections are ranked according to SNR, and the signal is reconstructed after eliminating the least informative components.

Starting with the connectivity matrix $\mathbf{W}^{\mathbf{spatial}}$ that represented the spatial network of the utilized sensor array (see Fig.5a), the graph Laplacian matrix $\mathbf{L}^{\mathbf{spatial}}$ was first derived. Based on the eigendecomposition of this matrix, $\mathbf{L}^{\mathbf{spatial}} = \mathbf{U}\,\mathbf{\Lambda}\,\mathbf{U}^{\mathbf{T}}$, and after placing the eigenvectors $\mathrm{U}_k$ in order of increasing eigenvalue $\lambda_k$, the matrix $\mathbf{U} = [\mathrm{U}_1, \mathrm{U}_2, \ldots, \mathrm{U}_{|V|}]$ was formed and used as the operator for performing GFT. The single-trial response signal $x_i^j(t)$, j=1,2,…$N_{sensor}$, i=1,2…,$N_{trial}$ was considered as a sequence $X_i(t)$ of momentary electric field distributions, or graph signals in sensor domain, which had been tabulated as columns in matrix $\mathbf{X_i}$. The corresponding sequence of projection within the GFT-domain was readily derived via a matrix operation $\mathbf{Z_i}=\mathbf{U}^{\mathbf{T}}\,\mathbf{X}_i$. We then treated the time-indexed projections along k$^{th}$ GFT-dimension $z_i^k(t)$, k=1,2,…$N_{component}$, i=1,2…,$N_{trial}$ as a multi-trial dataset and quantified the SNR of the evoked response [49], based on the post-stimulus interval of (50-250) milliseconds. Figure 9a includes the derived measurements and indicates that the stimulus induces “time-locking” only in some GFT-components (indicated via blue color). The spatial distribution of each one of these components is depicted topographically in Fig.9c. The shown distributions correspond to the spatial patterning of the selected GFT-eigenmodes and can be considered as spatial filters for the multichannel signal. The adopted procedure, therefore, quantified the contribution of the GFT-eigenmodes to the formation of multichannel evoked response. To further justify this, we provide in Fig.9b the result from trial-averaging the traces that corresponded to the single-trial projections from each of the selected eigenmode. Following this line of thought, we implemented a GFT-based filtering process by eliminating those components that were characterized by SNR lower than 1 and reconstructing (via inverse GFT) the multichannel signal only from the “reliable” GFT-components. This procedure actually realized a GSP-filter with a transfer function $h(\lambda_\kappa)$ indicated in Figure 9d and formulated in an elegant way as follows **Filtered_X$_\mathbf{i}$** = **H X$_\mathbf{i}$**(t) , with $\mathbf{H} = \mathbf{U}\,h(\boldsymbol{\Lambda})\,\mathbf{U}^{\mathbf{T}} = \sum_{\boldsymbol{k}} h(\lambda_\kappa)\ \mathrm{U}_k\,\mathrm{U}_k^{\mathrm{T}}$. The averaged multichannel signal after this filtering operation is included is Fig.9e using the same color-coding scheme with Fig.5. The enhanced visual response is associated with increased SNR (from 15% to 25% depending on the sensor), as can be seen by comparing the profiles in Fig.9f with the ones in Fig.5c.

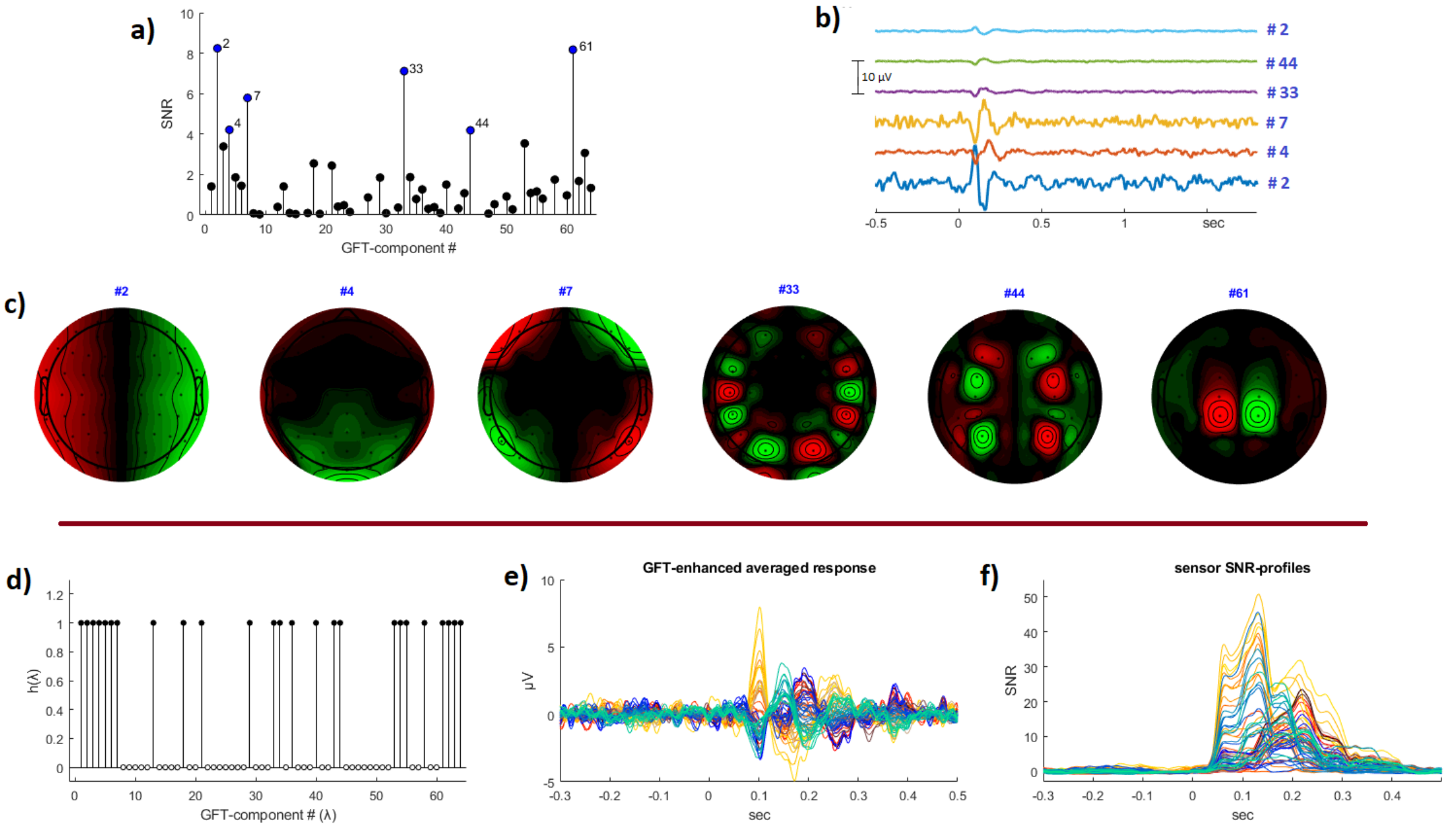

**Figure 9:** Graph filtering of multi-trial multichannel signal of visual evoked response.

## 3.5 A graph-theoretic comparison between passive and attentive responses.

Our section of worked out examples closes with a brief discussion on what has been learned by employing the previous graph-theoretic tools to contrast the attentive with the passive visual responses. Figure 10 includes both the ISOMAP average-response representation and the MST-based representation of single-trial variability for the attentive condition (in full analogy with Fig.5 and Fig.6 respectively). ISOMAP-derived graph shows a stronger stimulus-induced reorganization of the spatial network (PO7 appears as sharper tip). The MST-graph reveals a more uniform distribution and, accordingly, the selected single-trials (STs) are more similar with each other.

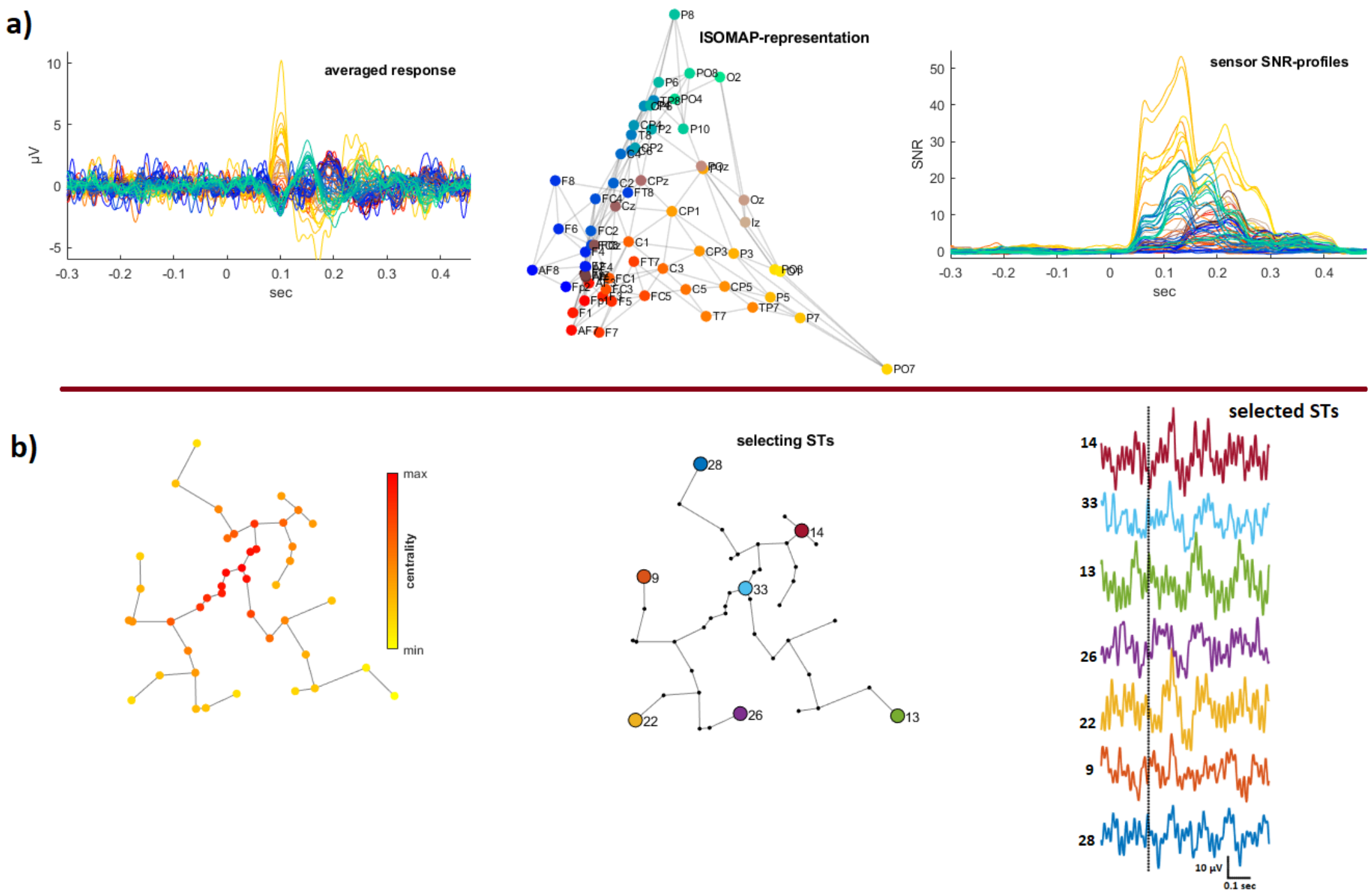

**Figure 10: a)** ISOMAP representation of attentive responses. **b)** MST-based representation of the trial-to-trial response variability at PO7 sensor.

A characteristic outcome from the joint analysis of the evolving, event-related, connectivity graphs of passive and attentive responses is presented in Fig.11. The included contrastive distance profile Profile(t)=dist(${}^{t}\mathbf{W}^{passive}$,${}^{t}\mathbf{W}^{attentive}$) (defined in line with the notation of section 1.2.3) was used to mine connectivity patterns at latencies where the response communication pattern was deviating the most between the two recording conditions. It is evident that attentive visual responses are associated with higher level of interconnectivity in β-rhythm functional network.

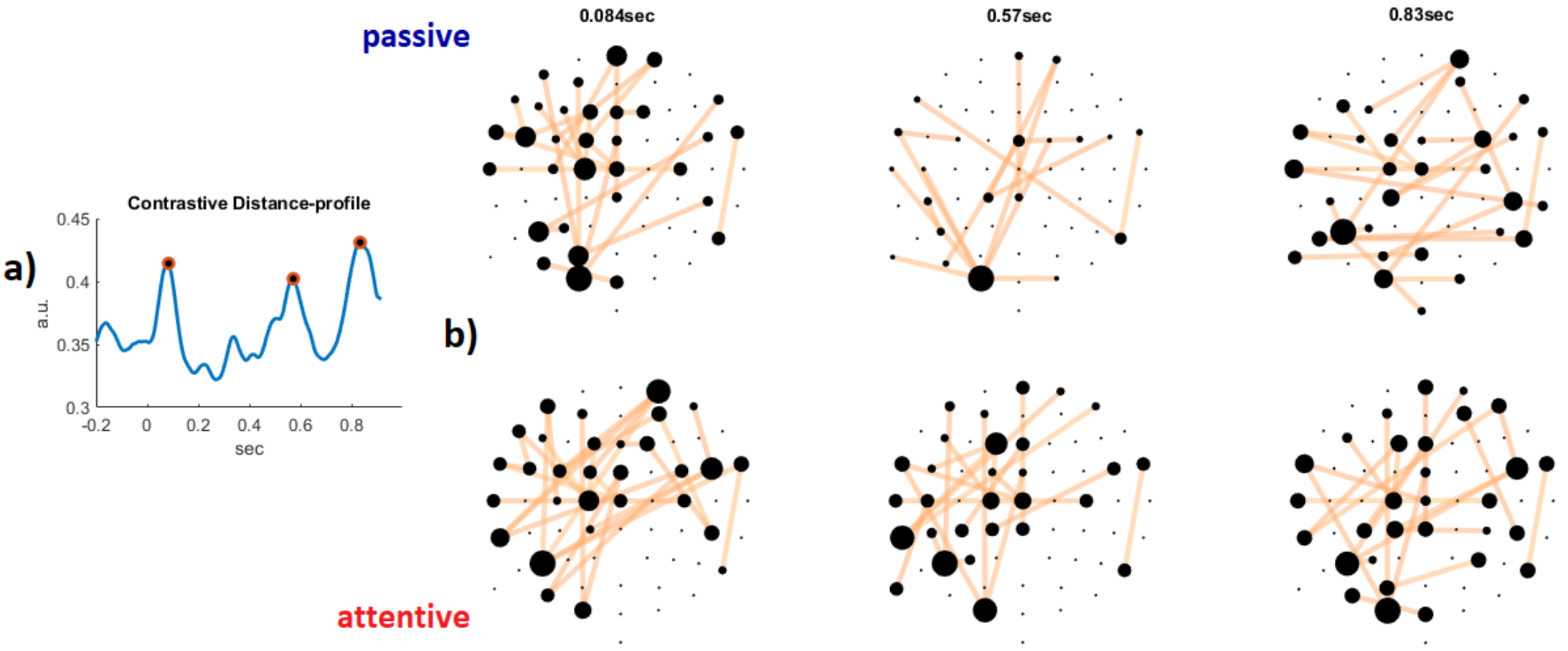

**Figure 11**: Mining snapshots of connectivity at the latencies where the evolution of β-rhythm functional network deviates mostly between the passive and attentive recording condition.

As a final example, we demonstrate the use of GFT-representation as a feature extraction step towards the classification of single-trial responses into passive and attentive ones. The temporal patterning of single-trial GFT projections $z^k_{attentive_i}(t)$ & $z^k_{passive_i}(t)$ within the early post-stimulus interval was analyzed, separately, for each component k=1,2,…$N_{component}$. Based on a non-parametric statistical test[4], we derived the WW-score [52], a class separability measure, as a means of feature ranking. Put simply, a GFT component that scores higher than 3 has enough pattern-analytic power to differentiate attentive from passive responses if it is treated as spatial filter and its output trace is fed to a classifier. Figure 12 includes the WW-score of each GFT-component and indicates the most informative ones. It is important to notice here that the discriminability in 4 GFT-projections is higher than the discriminability of the original sensors.

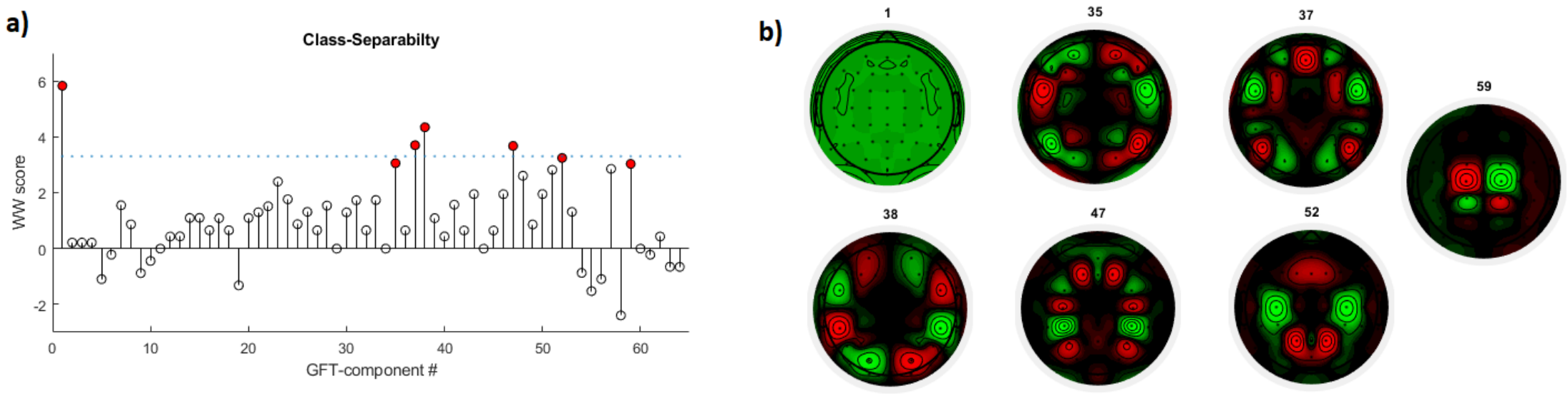

**Figure 12:** a) Pattern class-separability measure (WW-score) as a function of GFT-eigenmode (the blue dotted line indicates the maximal WW-score achieved by working at the individual sensor level). b) Selected GFT-eigenmodes corresponding to WW-score higher that 3.

[4] The Wald-Wolfowitz multivariate test, which has an inherent graph-theoretic character since it is based on MST.

# 4 DISCUSSION

Graph theory provides an optimal framework for handling neural interactions and shedding light on how coordinated brain activity supports cognition and ensures mental health. Recent theoretical innovations like hypergraphs [61] and multilayer / multiscale / multiplex / multislice networks [62, 63] are expected to initiate further fruitful explorations via graph-theoretic algorithms that operate beyond pairwise relations. We anticipate that the interplay between graph theory and brain signals will grow stronger in the years to come. The intriguing and closely related problem of learning graphs from signals [64], i.e. how to deduce the graph structure from the recorded brain activations, will provide alternative ways to measure connectivity. This will add to the known blending of graph theory and brain signal processing, which includes morphological descriptors that turn brain signals to graphs (such as the approach of *visibility graphs* [65]) and sophisticated descriptors of brain dynamics that derive connectivity-graphs (like the cross-frequency coupling estimators [22, 57]) and state-transition graphs [56].

In the course of this paper, the discussion on the utility of graph-theory in brain signal processing has been restricted, technically, to the analysis of event-related multichannel recordings. This by no means should leave the reader with the impression that the presented methodologies are strictly associated with the investigations for stimulus-induced brain's reaction and functional re-organization. It was a choice that motivated by the need for a simple unifying running example that could demonstrate the wide spectrum and power of graph-theoretic brain signal-analytics. Therefore, it should be underlined here that graph-theory also supports the research on brain's self-organization, as this is reflected in continuous-mode recordings from humans at rest [66, 67] or engaged in a naturalistic setting (like listening to music [35], driving a car [68], performing mental calculations [69], comprehending code [70] etc.). Actually, the network-characterization of functional connectivity at rest has been among the most popular research themes in human fMRI [17, 71] and EEG/MEG studies [72]. In some cases, whenever dynamics are considered, the timeseries of graphs may become overwhelming in terms of size. Then a parsimonious representation, by means of a clustering algorithm becomes necessary, which will either operate on time-resolved connectivity patterns or graphs. This naturally leads to the definition of connectivity/network microstates [56, 73, 74] for the description of brain's metastability [75].

In conclusion, graph theory and brain signal processing bind together inevitably in the endeavor to understand human brain function. The existing dialectic tensions are anticipated to be higher in the future, driven by the latest advances in deep-graph learning [76] and the spread of high-performance computing based on GPUs [77].